\documentclass[journal]{IEEEtran}

\ifCLASSINFOpdf
\else
\fi

\usepackage{graphicx}
\usepackage{color}
\usepackage{placeins}
\usepackage{float}
\usepackage{tabularx,colortbl}
\usepackage{gensymb}
\usepackage{cite}
\usepackage{amsmath,amssymb}
\usepackage{listings}
\usepackage{caption}
\usepackage{lipsum}
\usepackage{mwe}
\usepackage{breqn}
\usepackage{hhline}

\begin{document}

\title{Open-Loop Distributed Beamforming Using Scalable High Accuracy Localization}

\author{Sean M. Ellison,~\IEEEmembership{Student Member,~IEEE}, and Jeffrey A. Nanzer,~\IEEEmembership{Senior Member,~IEEE}% <-this % stops a space
\thanks{Manuscript received 2020.}
\thanks{This material is based in part upon work supported by The Defense Advanced Research Projects Agency under grant number N66001-17-1-4045, the National Science Foundation under grant number 1751655, and by the Air Force Research Laboratory under contract number FA8650-14-D-1725. The views, opinions, and/or findings contained in this article are those of the author and should not be interpreted as representing the official views or policies, either expressed or implied, of the Defense Advanced Research Projects Agency or the Department of Defense. \textit{(Corresponding Author: Jeffrey A. Nanzer)}}
\thanks{The authors are with the Department of Electrical and Computer Engineering, Michigan State University, East Lansing, MI 48824 USA (e-mails: elliso65@msu.edu, nanzer@msu.edu).}
}% <-this % stops a space

\markboth{IEEE}%
{Shell \MakeLowercase{\textit{et al.}}: Bare Demo of IEEEtran.cls for Journals}

\maketitle

\begin{abstract}
We present a distributed antenna array supporting open-loop distributed beamforming at 1.5 GHz. Based on a scalable, high-accuracy internode ranging technique, we demonstrate open-loop beamforming experiments using three transmitting nodes. To support distributed beamforming without feedback from the destination, 
the relative positions of the nodes in the distributed array must be known with accuracies below $\lambda/15$ of the beamforming carrier frequency to ensure that the array maintains at least 90\% coherent beamforming gain at the receive location. For operations in the microwave range, this leads to range estimation accuracies of centimeters or lower. We present a scalable, high-accuracy waveform and new approaches to refine range measurements to significantly improve the estimation accuracy. Using this waveform with a three-node array, we demonstrate high-accuracy ranging simultaneously between multiple nodes, from which phase corrections on two secondary nodes are implemented to maintain beamforming with the primary node, thereby supporting open-loop distributed beamforming. Upon movement of the nodes, the range estimation is used to dynamically update the phase correction, maintaining beamforming as the nodes move. We show the first open-loop distributed beamforming at 1.5 GHz with two-node and three-node arrays, demonstrating the ability to implement and maintain phase-based beamforming without feedback from the destination.

%These localization requirements are met by a specially sparse, two-tone stepped frequency waveform which provides a range estimation through time delay information to a reference point in space, in this case a primary node. The range estimation is converted to a phase shift of operational frequency and applied to all subsequent secondary nodes, allowing for operations where no feedback from target location is available and reducing the effects of array dynamics.
% Operations containing a primary node and up to two secondary nodes simultaneously performing their relative phase adjustments are shown.
\end{abstract}


\IEEEpeerreviewmaketitle

\section{Introduction}
Progress in wireless communications systems, as well as other wireless applications such as remote sensing, radar, and imaging, depends on the ability to continually improve the capabilities of antennas, arrays, and transceivers in terms of power, gain, throughput, and resolution, among other metrics. Improvements in these metrics are often achieved by focusing on new individual components and subsystems, resulting in better individual platform-based systems. However, limitations of this platform-centric model, stemming from constraints on size, weight, cost, and power consumption, make it increasingly challenging to scale such performance metrics on a single platform. To address this challenge, recent research has focused on the development of distributed wireless technologies, where collections of small, relatively inexpensive wireless systems are coordinated to mimic the performance of a single, large system, or to achieve performance otherwise unattainable with a single platform. Implemented in Multiple Input Multiple Output (MIMO)~\cite{7128345, 1353475, 1589439,6352830} or distributed beamforming~\cite{7880558,4202181,4588338,7462174} applications, such distributed wireless systems enable direct performance scalability by adding or removing inexpensive nodes from the array \cite{8471118,6657792}.

Coherent distributed arrays (CDAs) are a particular form of distributed wireless system where individual wireless elements coordinate at the level of the radio frequency (RF) phase to enable distributed beamforming \cite{4202181,6779684,7803582}. Coordination of separate moving nodes is a challenging problem, in which the following three fundamental coordination tasks must be accomplished: frequency synchronization to ensure all elements are operating at the same reference frequency \cite{9028079,1408270,5482522,8889331}; time alignment to ensure that there is sufficient overlap of the information at the target destination \cite{5893884,8378649,chatterjee2019}; and phase alignment to enable constructive interference at the target. Phase alignment presents the most challenges due to the extremely small tolerance to errors at microwave frequencies. Past works approached this issue through closed-loop feedback loops where a receiver co-located with the target location provides information to the transmitting array, from which the transmitters can determine how to adjust their relative phases to converge to a phase-coherent state at the receiver. Among these closed-loop approaches are receiver-coordinated explicit-feedback \cite{6488994}, three-bit feedback \cite{5670903}, primary-secondary synchronization \cite{4202181}, reciprocity where channel estimation is performed from signals sent by the target receiver \cite{7460546}, round-trip synchronization \cite{4542555}, and two-way synchronization \cite{5957340}. Although these feedback methods are effective, they are restricted to applications where reliable feedback can be provided by the destination. Numerous situations arise where such feedback is not available, particularly in cases where individual nodes in the array do not have sufficient sensitivity to close a link to a base station on their own. Furthermore, closed-loop architectures are inherently unable to support wireless applications beyond communications, such as remote sensing, imaging, and radar where coherent feedback is generally not present.

Open-loop CDAs (i.e. feedback-free distributed arrays) are types of CDAs where feedback from the destination location is not leveraged, thus the array self-coordinates to a phase-aligned state \cite{7803582,7782842}. To achieve phase alignment and implement a phase-based beamsteering operation, the relative positions of the individual nodes must be known to within a fraction of the wavelength of the beamforming frequency. Previous works have shown that these internode ranging measurements must have accuracies of less than $\lambda/15$ to have no more than 0.5 dB reduction in coherent gain with a probability of 90\% \cite{7803582,8870258}. To operate in the microwave range this calls for accuracies on a sub-centimeter level. Furthermore, this accuracy must be obtained before nodes move out of coherence due to array dynamics. This has been achieved in the past using optical systems but such systems are not easily scalable and require accurate tracking and pointing for each node connection \cite{4026149,189658}. We have previously shown that spectrally-sparse waveforms can achieve near-optimal ranging accuracy, and can be designed to support simultaneous ranging between multiple nodes, while using a microwave link that does not require accurate pointing and tracking \cite{9057428}. This method of phase adjustment through localization for open-loop arrays presents a much more general approach  than that of closed-loop systems, enabling applicability to situations where array gain is necessary to initialize the link and to sensing applications where coherent feedback is not available.

In this work, we present what is, to the best of our knowledge, the first open-loop distributed beamforming system using a scalable, high-accuracy ranging waveform with three transmitting nodes. The ranging waveform is based on a two-tone stepped-frequency waveform that supports ranging to multiple nodes simultaneously. We demonstrate the use of this waveform in an open-loop distributed transmitter array consisting of three separate 1.5 GHz transmitters. One node is designated as the primary node, with the two additional secondary nodes performing high-accuracy ranging from which the secondary nodes implement a phase-coherent transmission to cohere at a separate receiver. We demonstrate coherent signal summation at end-fire orientation that is maximally impacted by range-induced phase errors and thus represents the most challenging beamforming case.
%the first open-loop system to simultaneously perform dynamic phase alignment to enable coherent microwave beamforming is presented. The TTSFW is used for localization with a total operational bandwidth of 12.5 MHz to enable an experimental operational frequency of 1.5 GHz with dynamic phase adjustments form ranging data. Experiments are conducted utilizing a primary node, to set a phase reference in space, and one to two secondary nodes which adjust their output to the primary reference to ensure constructive interference at the target. 
In Section \ref{Sec1} we describe the waveform parameters associated with inter-node localization followed by Section \ref{Sec2} where the hardware setup associated with both the primary and secondary nodes is discussed. Section~\ref{Sec3} reviews the tolerance of gain degradation due to errors in position  and explores the inherent uncertainty that is present in the localization waveform. Methods for improvement in localization, such as retransmit gain from the primary, matched filtering, interpolation, and Kalman filtering, and are discussed in Section \ref{Sec4} followed by a discussion of the measurements incorporating dynamic phase shifting to implement distributed beamforming.

%%%%%%%%%%%%%%%%%%%%%%%%%%%%%%%%%%%%%%%%%%%%%%%%%%%%%%%%%%%%%%%%%%%%%%%%%
\section{Range Estimation Waveform}\label{Sec1}

\begin{figure}[t!]
\begin{center}
\noindent
  \includegraphics[width=\linewidth]{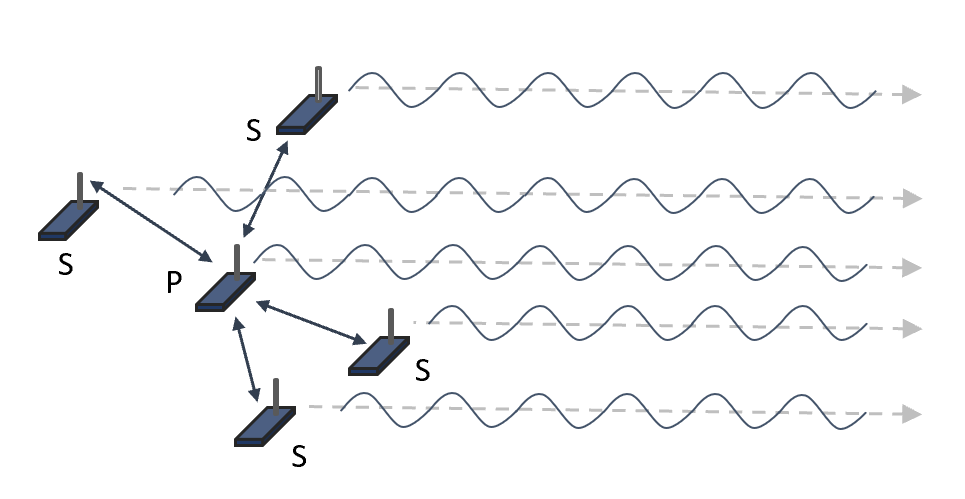}
  \caption{Distributed antenna array topology using a primary-secondary ranging approach.}
	\label{fig:OV1}
\end{center}
\end{figure}

The array design in this work is based on a primary-secondary internode ranging architecture. A diagram of this type of array design can be seen in Fig. \ref{fig:OV1}. Because the relative ranges between nodes is of importance, each node in the array must perform a ranging measurement between another node such that the array forms a connected graph. Each node can then implement a beamforming operation in a pair-wise manner \cite{7803582}. After estimating the baseline $d$ between the two nodes, the phase at the node performing the measurement is updated by 
\begin{equation}\label{phase}
\phi_s = 2\pi f \tfrac{d}{\lambda}\cos\theta,
\end{equation}
where $f$ is the beamforming carrier frequency, $\lambda = \tfrac{c}{f}$ is the wavelength with $c$ the speed of light, and $\theta$ is the steering angle representing an additional source of error which is maximized during end-fire operations. % The latter term represents an additional source of error in the beamforming process, and as \eqref{phase} shows, when the two-element pair is beamsteering towards end-fire ($\theta = 0^{\degree} / 180^{\degree}$) the error in range produces the maximum phase error. 
In this work we demonstrate coordination in a linear array; extensibility to general array layouts further requires estimation of the angle of each secondary node to the primary. Angle-of-arrival (AOA) methods have been explored in literature \cite{alsubaie2014multiple}, the most notable of which are MUSIC \cite{1143830} and ESPRIT \cite{roy1989esprit}. We focus exclusively on the ranging challenge in this work, however additional error terms that arise from angle estimation along with beamforming to arbitrary directions can be included in a total error budget \cite{7803582}. 
%The impact of systematic error terms is analyzed in detail in Section \ref{Sec2}.

To address the internode ranging (i.e. the term $d$ in \eqref{phase}), we previously investigated the use of spectrally-sparse waveforms for high-accuracy ranging, showing that a two-tone waveform obtains near-optimal ranging accuracy \cite{7801084,8930595}. Addressing the ambiguity challenges with a simple two-tone waveform, we previously investigated  scalability approaches by developing a two-tone stepped-frequency waveform (TTSFW) \cite{9057428}, however it was found that the range estimates from direct matched-filtering of the ranging waveform varied upon relatively fast movement of the nodes with a magnitude too great for reliable beamforming. This work expands on prior work by implementing a Kalman filter to improve the robustness of the range measurement, and demonstrating the ability to maintain a steered beam when the nodes are moved.

%Note that this method of updating beamforming phase based only on a radial distance estimation will limit the feasible array implementation to a single dimension since only a single baseline can be measured. 

The TTSFW waveform consists of a two-tone waveform that is pulse-modulated, with a change in carrier frequency in each successive pulse. The baseband waveform can be written as
\begin{equation}\label{eq:1}
	s(t)=\frac{1}{N}\sum_{n=0}^{N-1}\text{rect}\left(\frac{t-nT_r}{T}\right)\left(e^{j2\pi f_1t}+e^{j2\pi f_2t}\right)e^{j2\pi n \delta f t}
\end{equation}
where $f_1$ is the lower of the two tones per pulse, $f_2$ is the upper tone, $\delta f$ is the frequency step, $N$ is the number of pulses, $T$ is the waveform duration, and $T_r$ is the active portion of the pulse (i.e. $T$ multiplied by the duty cycle). The frequency step can be given by
\begin{equation}
	\delta f=\frac{BW}{2N-1}
\end{equation}
where $BW$ is the total waveform bandwidth. The upper frequency in each two-tone pulse is then given by $f_2=f_1+\Delta f$ where
\begin{equation}
	\Delta f=N\delta f
\end{equation}
The waveform is generated such that $N!$ is greater than or equal to the number unique connections between the primary and all secondary nodes in the array. This ensures that every secondary node has a unique pulse signature, which allows for simultaneous measurements utilizing the same bandwidth with minimal interference. This baseband waveform is then upconverted to a carrier frequency $f_c$, which is in general different from the beamforming carrier frequency $f$.

\begin{figure} [t!]
	\centering
\includegraphics[width=\linewidth]{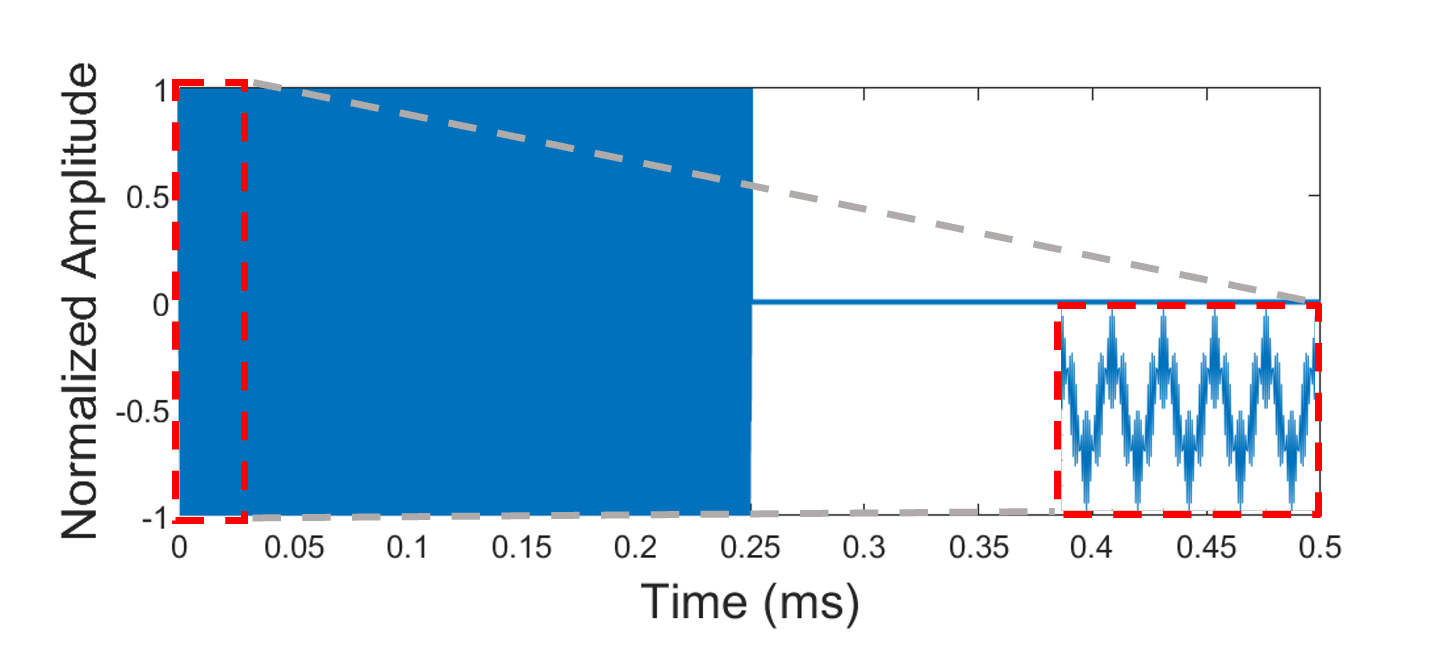}

(a)

\includegraphics[width=\linewidth]{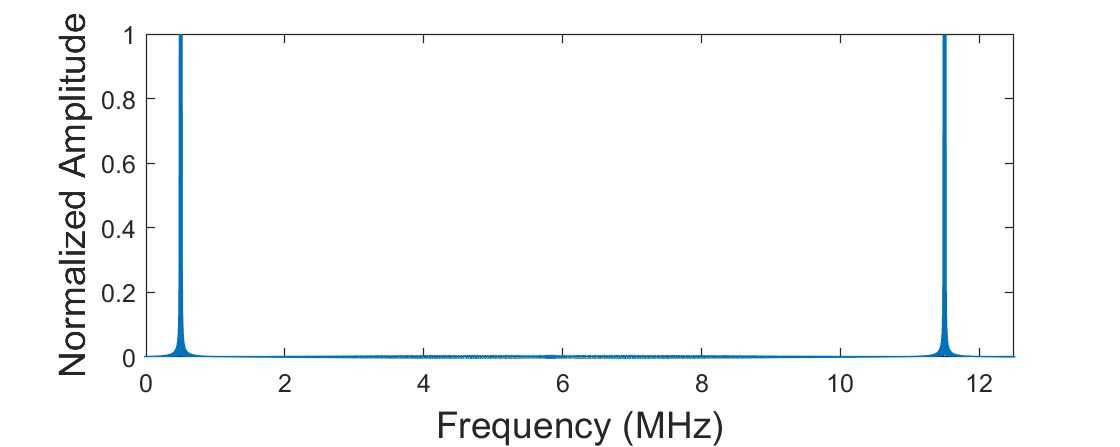}

(b)

	\caption{Measured waveform supporting ranging between two nodes: (a) time domain; (b) frequency domain.}
	\label{fig:waveform}
\end{figure}

In the case of a two-node array, consisting of a primary node and a single secondary node, only a single pulse $N=1$ is required since only a single unique connection is made. Fig. \ref{fig:waveform} shows an example implementation of this waveform in the time and frequency domains. For this case the baseband waveform had a duration of 1 ms with a 50\% duty cycle and frequencies $f_1 = 500$ kHz and $f_2 = 11.5$ MHz, which matches the bandwidth that has been demonstrated to obtain high ranging accuracy \cite{9057428}.

\begin{figure} [t!]
	\centering
\includegraphics[width=\linewidth]{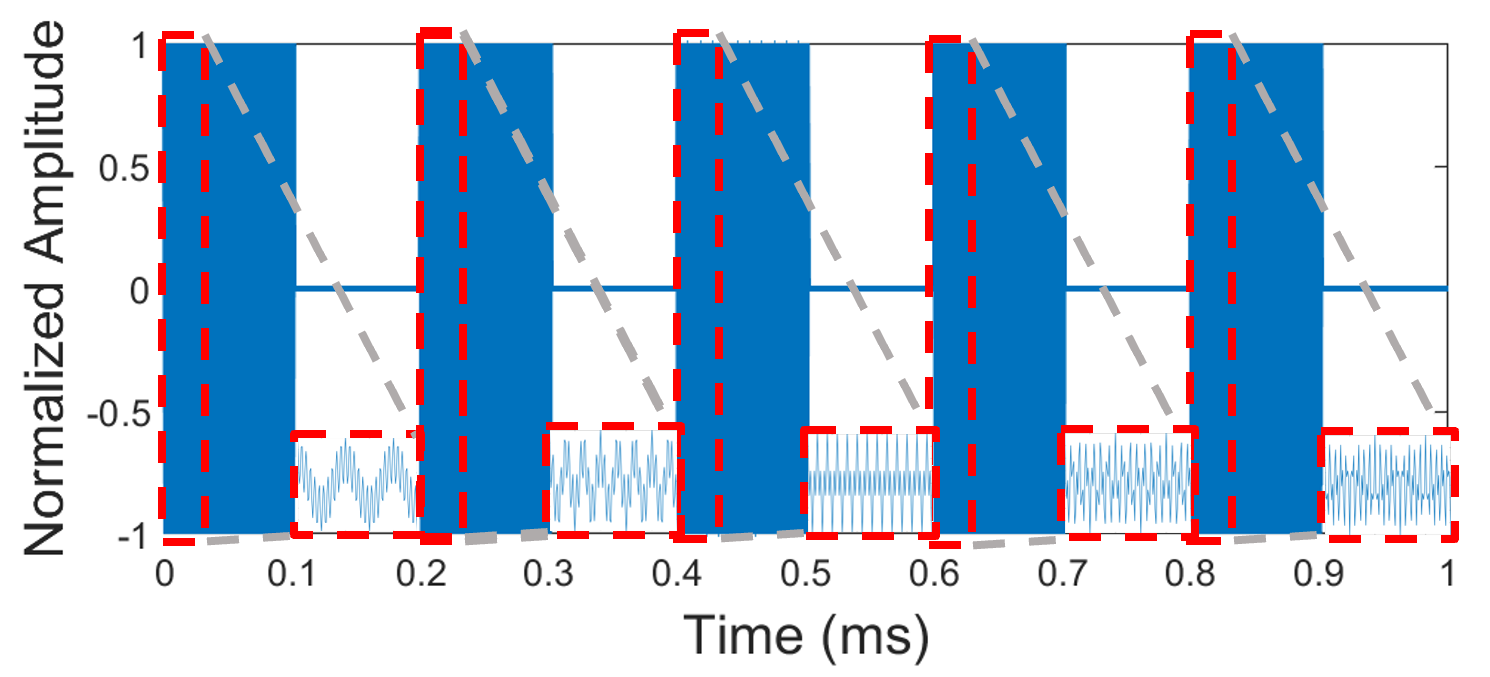}

(a)

\includegraphics[width=\linewidth]{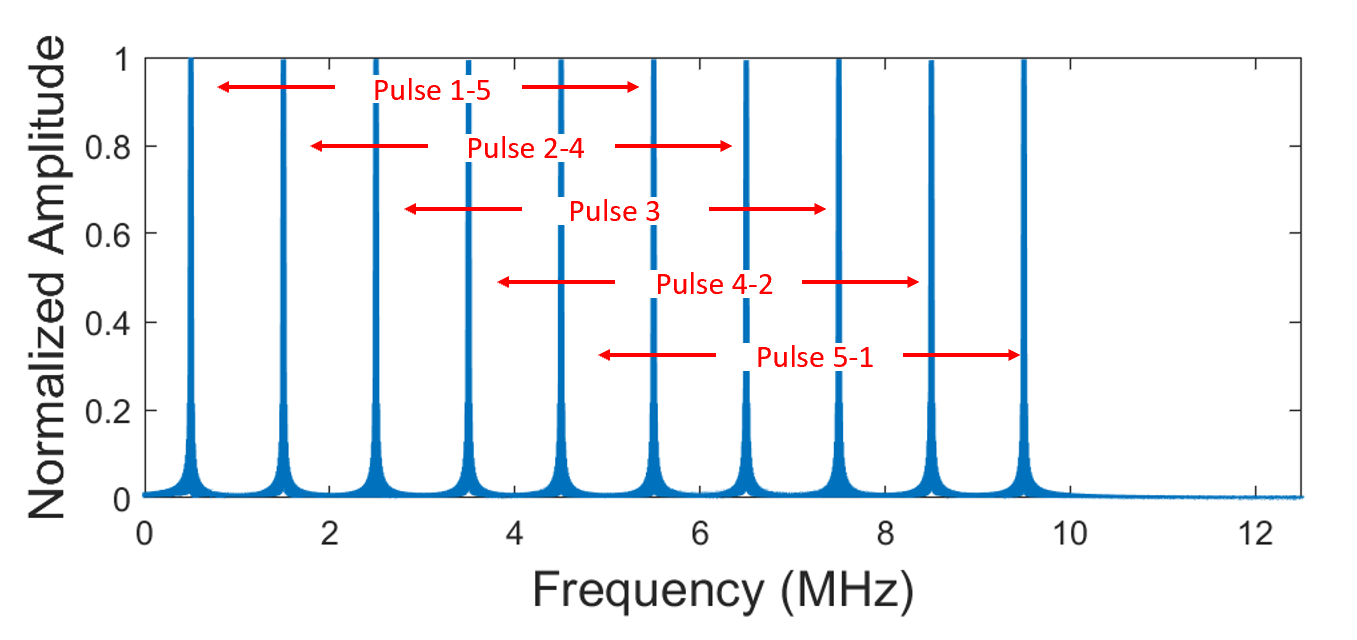}

(b)

	\caption{Measured waveform supporting ranging between five nodes: (a) time domain; (b) frequency domain.}
	\label{fig:waveform2}
\end{figure}

In the case of a three-node array, consisting of a primary node and two secondary nodes, there are two unique connections and therefore a multipulse system is required such that $N\geq2$. We demonstrate here a waveform supporting more than three nodes, with $N=5$ to illustrate the scalability. An image of the waveform in the time domain and frequency domain can be seen in Fig. \ref{fig:waveform2}. The time duration of the baseband waveform was 200 $\mu$s per pulse with a 50\% duty cycle, frequencies $f_1 = 500$ kHz and $f_2 = 5.5$ MHz, and a step sized of $\delta f=1$ MHz. To ensure that each connection has a unique pulse signature, one secondary node begins its waveform with the pulse containing the lowest frequency pair of tones and increased the frequency by the frequency step, while the second secondary node begins with the pulse with the highest frequency pair and decreased the frequency by the frequency step; this approach to unique pulse-to-pulse signatures per node is easily extendable.  The pulse labels in Fig. \ref{fig:waveform2}(b) given by Pulse $a-b$ indicate the pulse order relative to the secondary such that the index $a$ is associated with secondary one while $b$ is associated with secondary two. 

%%%%%%%%%%%%%%%%%%%%%%%%%%%%%%%%%%%%%%%%%%%%%%%%%%%%%%%%%%%
\section{Ranging Requirements for Open-Loop Distributed Beamforming}\label{Sec2}
%\subsection{Positional Error Tolerance}
For a distributed array of $N$ elements that are arbitrarily positioned in space and continuously transmitting, the far-field response is given by
\begin{equation}\label{eq:2}
	s_r(t)=\sum_{n=1}^Nh_n\alpha_n(t)e^{j\left(2\pi f t+\phi_n+\phi_{s,n}\right)}
\end{equation}
where $N$ is the number of array nodes, $h_n$ is the channel response between node $n$ and the target, $\alpha_n$ is the amplitude, $\phi_{s,n}$ is the delay imparted at node $n$ by the node separation, given by \eqref{phase}, and $\phi_n$ is the phase correction with errors given by
\begin{equation}
	\phi_n=\frac{2\pi}{\lambda}(d_n+\delta d_n)\cos\theta_n
\end{equation}
where $\delta d_n$ is error in localization, $\theta_n$ is the error in beamsteering angle, and $\delta \phi_c$ is the relative phase error of the internal oscillators. Here we consider the most challenging case in terms of ranging errors with end-fire beamforming, where $\theta_n = 0$ and thus $\phi_n = \tfrac{2\pi}{\lambda}(d_n+\delta d_n)$.

\begin{figure}[t!]
\begin{center}
\noindent
  \includegraphics[width=\linewidth]{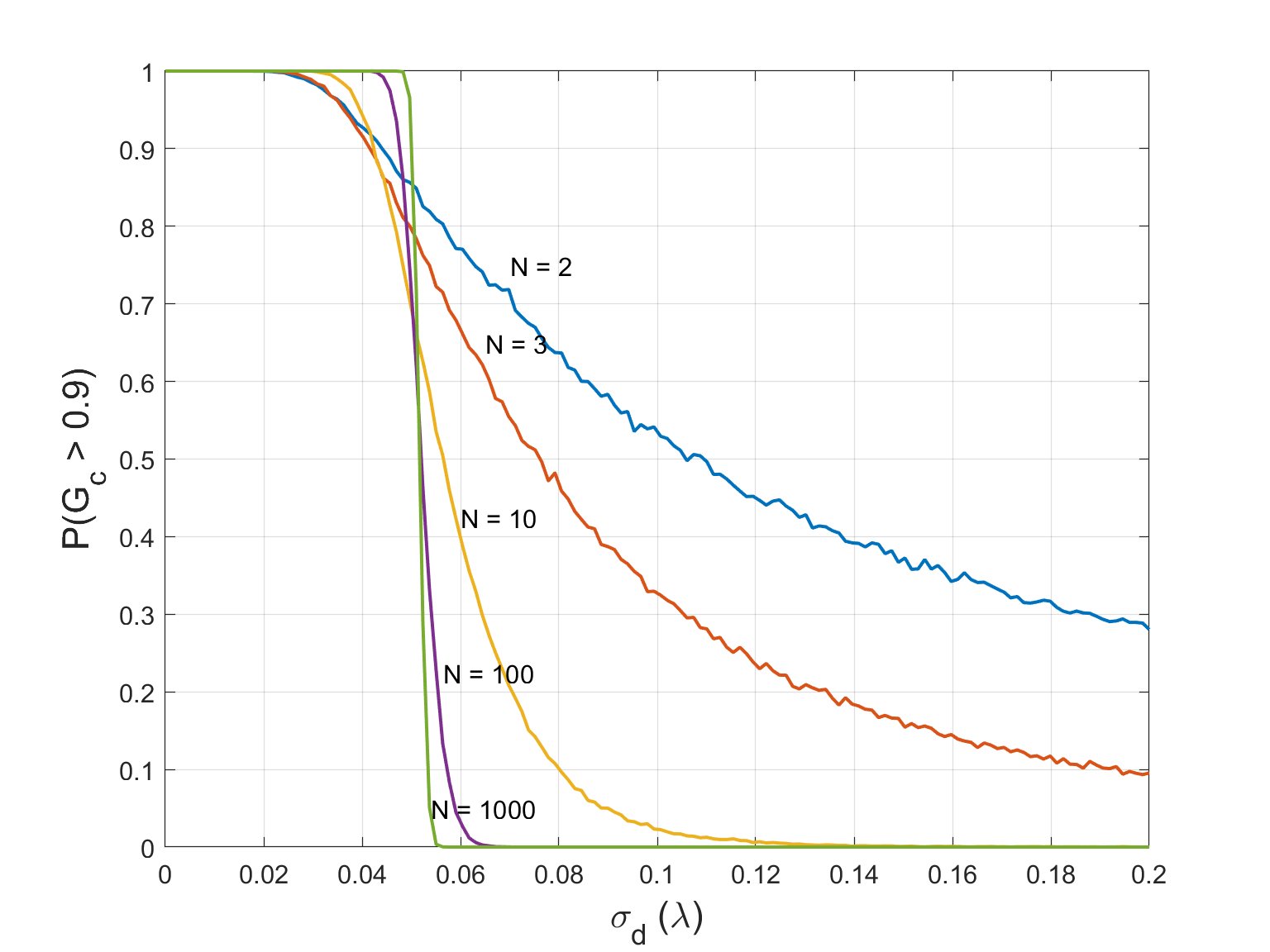}
	\caption{Probability of obtaining coherent gain above 90\% ($P(G_C\geq 0.9)$) for various array sizes versus standard deviation of internode range error for the special case of end-fire configuration. As array becomes large, the threshold for $P(G_C\geq 0.9) = 1$ approaches $\lambda/20$.}
	\label{Gc}
\end{center}
\end{figure}

A perfectly aligned system would produce an ideal signal given by
\begin{equation}\label{eq:3}
	s_i(t)=C\sum_{n=1}^{N}e^{j\left(2\pi ft-\frac{2\pi}{\lambda}d_n\right)}
\end{equation} 
To evaluate the effects of localization errors on the power of the beamformed signal, the power of the signal with errors (\ref{eq:2}) is evaluated relative to that of the ideal signal (\ref{eq:3}) by
\begin{equation}
	G_c=\frac{|s_rs_r^*|}{|s_is_i^*|}
\end{equation}
The standard deviation of the error term $\sigma_d$ was varied and the probability of exceeding a particular threshold $P(G_C\geq X)$, where $0\leq X\leq1$, was determined through 100,000 Monte-Carlo simulations. In this work we consider the coherent gain to be greater than 90\% of that of the ideal array ($X=0.9$). We previously showed that as the array size increases, obtaining $P(G_C\geq X)$ requires ranging errors of $\lambda/15$ or less for arbitrary steering angles \cite{7803582,8870258}.  For the more stringent end-fire case, this requirement is closer to $\lambda/20$, as shown in Fig. \ref{Gc}, where the result of 10,000 Monte Carlo simulations is shown for $P(G_C\geq 0.9)$. For microwave and millimeter-wave beamforming frequencies, this requirement leads to ranging accuracies on the order of $cm$ or $mm$.
%\begin{figure}[t!]
%\begin{center}
%\noindent
  %\includegraphics[width=\linewidth]{pos_error.png}
  %\caption{Probability of coherent gain in the presence of positional errors. The threshold for $P(G_c\geq0.9)=1$ approaches $\lambda/15$ (from \cite{7803582})}
	%\label{fig:error}
%\end{center}
%\end{figure}

%\subsection{Waveform Uncertainty}
Inherent uncertainly is present in any system that attempts to estimate a parameter due to the presence of random noise. A measure of the variance or stability of an estimate of a random variable is given by the Cramer-Rao Lower Bound (CRLB), which for time delay estimation is given by \cite{7801084}
\begin{equation}\label{eq:4}
	\text{var}(\hat{\tau}-\tau)\geq\frac{1}{\zeta_f^2\text{SNR}}
\end{equation}
where $\zeta_f^2$ is the mean-squared bandwidth of the waveform and SNR is the signal-to-noise ratio. This can be converted to the two-way range estimate by
\begin{equation}\label{crbrange}
	\text{var}(\hat{x}-x)\geq\frac{c^2}{4\zeta_f^2\text{SNR}}
\end{equation}
where the factor of 4 derives from the two-way propagation seen by typical radar measurements. Thus, \eqref{crbrange} gives a measure of the minimum amount of positional variability that can be obtained for a given waveform. For a given SNR, the achievable variance is directly dependent on the mean-squared bandwidth, given by \cite{7801084}
\begin{equation}
	\zeta_f^2=\frac{\int_{-\infty}^{\infty}(2\pi f)^2|S(f)|^2df}{\int_{-\infty}^{\infty}|S(f)|^2df}
\end{equation}
where $S(f)$ is the spectrum content of the waveform. This can be derived for the TTSFW as \cite{9057428}
\begin{equation}
	\zeta_f^2=\pi^2\left(\frac{BW}{2-\frac{1}{N}}\right)^2+\frac{(2\pi BW)^2}{N(4N^2+4N+1)}\sum_{n=0}^{N-1}n^2
\end{equation}
For a given number of pulses $N$, $\zeta_f^2$ is maximized when the bandwidth $BW$ is also maximized and thus providing the minimum variance. This is the driving motivation behind spectrally sparse waveforms as a method for localization. The mean-square bandwidth for the two waveforms described in Section \ref{Sec1} can be derived as
\begin{align*}
	\zeta_{f,\text{2 node}}^2\Big|_{N=1}&=\pi^2 BW^2=1.942\times10^{15}\\
	\zeta_{f,\text{3 node}}^2\Big|_{N=5}&=\frac{507\pi^2 BW^2}{1000}=6.0547\times10^{14}
\end{align*}

We evaluate the lower bound on delay estimation for the two waveforms considering an SNR of 30 dB; this closely matches that obtained in typical cooperative ranging measurements in distributed arrays. We note that, unlike a traditional radar ranging measurement which undergoes propagation losses in both directions, the cooperative ranging implemented in a distributed antenna array only suffers losses in one direction, with the primary node repeating the signal with gain. Thus, relatively high SNR values are feasible. 
%The  hardware implementation is described in more in-depth in Section \ref{Sec4}.
%Both of the two and three node measurements are taken at approximately 30 dB preprocessing SNR which was determined by eigenvalue decomposition \cite{6860836,8870258}. 
The processing gain resulting from the matched filter process, a method for time delay estimation, is equivalent to the time-bandwidth product of 
$NT_rBW_n$,
where $T_r$ is the non-zero time duration of the pulse, $N$ is the number of pulses, and $BW_n$ noise bandwidth which for this work, since no additional filtering outside the analog bandwidth of the system is used, is equal to the sampling bandwidth of 12.5 MHz.

For the two node experiment, $T_r = 250$ $\mu$s, thus the processing gain was 35 dB resulting in a post processing SNR of 65 dB. Using (\ref{eq:4}) the bound on the variance of the time delay can be derived as $\sigma_\tau^2=2.65\times10^{-22}$ $s^2$ which can be converted to the two-way distance variability of $\sigma_x=2.44$ mm. The distance uncertainty sets the maximum operation frequency that can be achieved by
\begin{equation}
	f\leq\frac{c}{20\sigma_x}
\end{equation}
where $\sigma_x$ is the standard deviation of the two way distance measurement and the factor of 20 derives from the coherent gain statistical analysis for the end-fire array configuration. For this case the resulting maximum frequency limit is $f_{\text{2 node}}\leq6.14$ GHz.

For the three node experiment, $NT_r$ resulting from the summation of all the pulses was equal to 500 $\mu$s, yielding a processing gain of 38 dB resulting in a post processing SNR of 68 dB. The time delay variance was thus $\sigma_\tau^2=2.62\times10^{-22}$ $s^2$, and the distance variance $\sigma_x=2.42$ mm. The maximum operational frequency for the three node is thus derived to be $f_{\text{3 node}}\leq6.18$ GHz. 
%To make sure to stay well below this limit due to added measurement uncertainty, the coherent action is arbitrarily chosen to be a 1.5 GHz sinusoid for both experiments.

%%%%%%%%%%%%%%%%%%%%%%%%%%%%%%%%%%%%%%%%%%%%%%%%%%%%%%%%%%%%%%%%
\section{Range Estimation Refinement for Node Motion}\label{Sec3}
Phase alignment of the operational frequencies to produce a phase coherent signal at the target destination is done by estimating the range between the primary node and the corresponding secondary node and applying the appropriate phase shift. For this experiment, the ranging signal is transmitted from the secondary node(s) to the primary node where a repeater captures the signals and retransmits. The center frequencies of the transmit and receive of the repeater are separated by 1 GHz. By doing this the propagation losses are proportional to $1/R^2$ rather than $1/R^4$ which is seen by typical radar measurements. This will also ensure that the desired signal will dominate any multipath and any crosstalk can be neglected. After the signal is received, the secondary node estimates the time of flight by matched filtering the return signal. The peak of the matched filter is spline interpolated in real-time in LabView with a thousand points to improve the accuracy. 
%This interpolation is done using a built-in LabView spline function and is limited to a thousand because any higher would result in system failure due to limitations on LabView processing speeds. 

\begin{figure}[t!]
\begin{center}
\noindent
  \includegraphics[width=\linewidth]{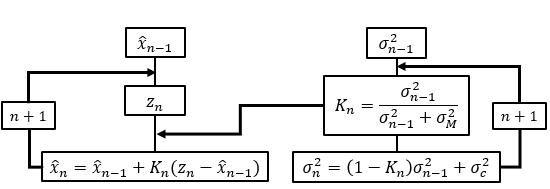}
	\caption{Block diagram of the Kalman filter used for range estimation refinement with moving nodes.}
	\label{kalman}
\end{center}
\end{figure}

The matched filter output peak value is tracked using a 1-D Kalman filter. A Kalman filter gives the optimal state estimation for linear systems in the presence of Gaussian noise \cite{kalman1960new}. The model of the filter, shown in Fig. \ref{kalman}, is 
\begin{equation}
	\hat{x}_n=\hat{x}_{n-1}+K_n(z_n-\hat{x}_{n-1})
\end{equation}
where $\hat{x}_n$ is the prediction of the current state, $\hat{x}_{n-1}$ is the prediction of the previous state, $z_n$ is the measurement at the current state, and $K_n$ is the current state Kalman gain given by
\begin{equation}
	K_n=\frac{\sigma^2_{n-1}}{\sigma^2_{n-1}+\sigma^2_M}
\end{equation}
where $\sigma^2_{n-1}$ is the previous state uncertainty and the measurement variance $\sigma^2_M=3\times10^{-5}$ is determined by measuring the variance of 1000 peak values. The current state uncertainty is then updated by
\begin{equation}
	\sigma^2_{n}=(1-K)\sigma^2_{n-1}+\sigma^2_c
\end{equation}
where an additional constant uncertainty of $\sigma_c^2=5\times10^{-6}$ is added to model the array dynamics seen in the matched filter output. This is to model the internode range of the system as a constant value with a small, random perturbation to account for both positive and negative radial motion at random time instances. The experiments in the following section have induced motion that is proportional to the wavelength of the 1.5 GHz coherent frequency (20 cm), which is roughly 2\% of the sampling interval of the matched filter at 25 MHz (12~m). Since this induced motion is very small compared to the discretization of the matched filter, this acts like a fluctuation on an otherwise constant value and therefore a  1-D Kalman filter is sufficient. However, if the motion is much larger than a fraction of the matched filter sampling rate, resulting in discontinuities in the matched filter output, the Kalman filter will diverge. If this is the case other techniques such as a higher dimensional Kalman filter to measure velocity and acceleration, such as an extended Kalman filter (EKF) \cite{1271397,jazwinski2007stochastic,sorenson1985kalman} to linearize the discontinuities or an unscented Kalman filter (UKF) \cite{julier1997new,1025369,wan2000unscented}, may be needed.
 %but is unnecessary of this experiment. 
The time delay estimate found from the output of the Kalman filter is converted to a phase shift of the operational frequency. This phase shift is then applied to the beamforming carrier signal on the secondary node. Thus, as the primary and secondary change their relative positions, the outputs remain phase locked at the target location.

%%%%%%%%%%%%%%%%%%%%%%%%%%%%%%%%%%%%%%%%%%%%%%%%%%%%
\section{Distributed Antenna Array and Open-Loop Distributed Beamforming Experiments} \label{Sec4}

\begin{figure} [t!]
	\centering
\includegraphics[width=\linewidth]{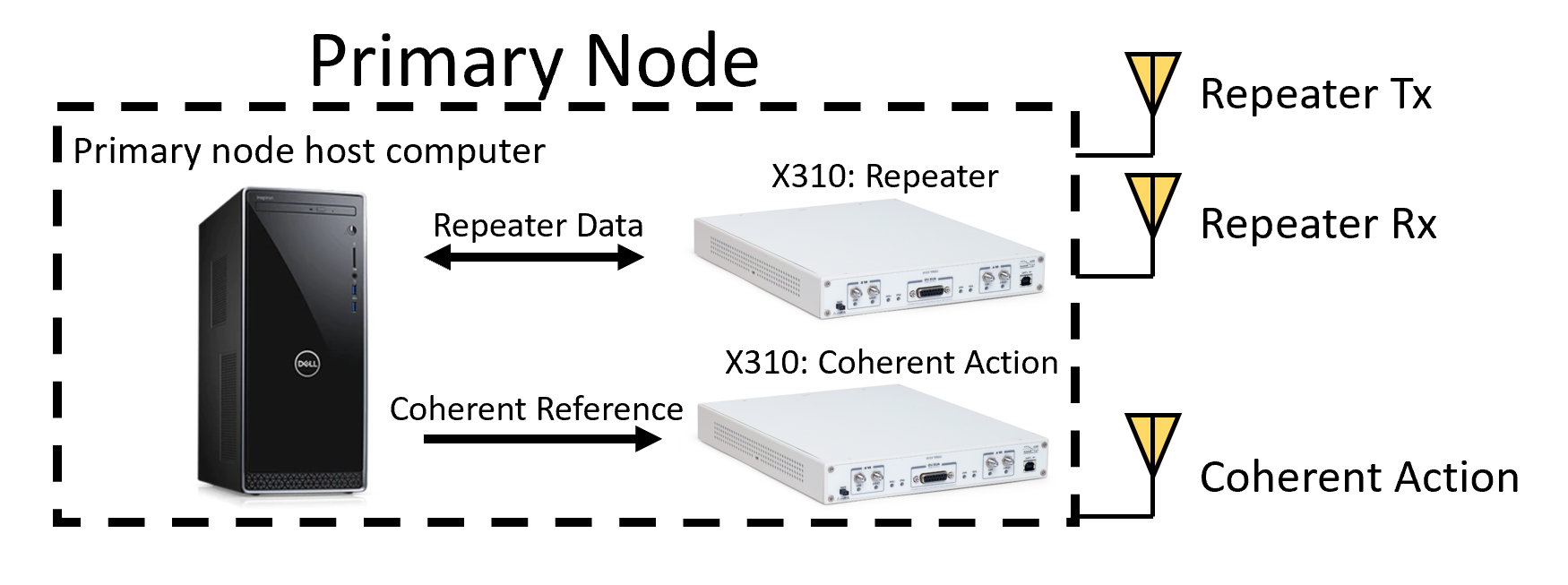}

\includegraphics[width=\linewidth]{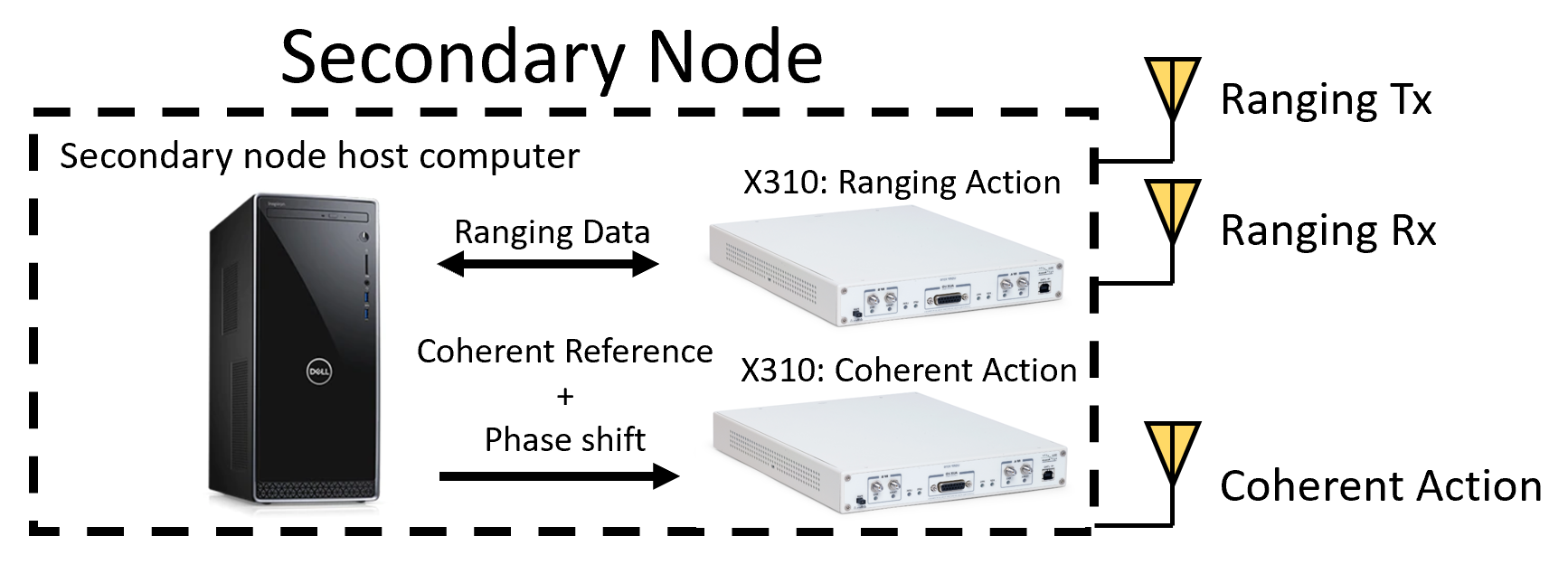}

	\caption{Block diagrams of the nodes.}
	\label{fig:nodes}
\end{figure}

\begin{figure}[t!]
\begin{center}
\noindent
  \includegraphics[width=\linewidth]{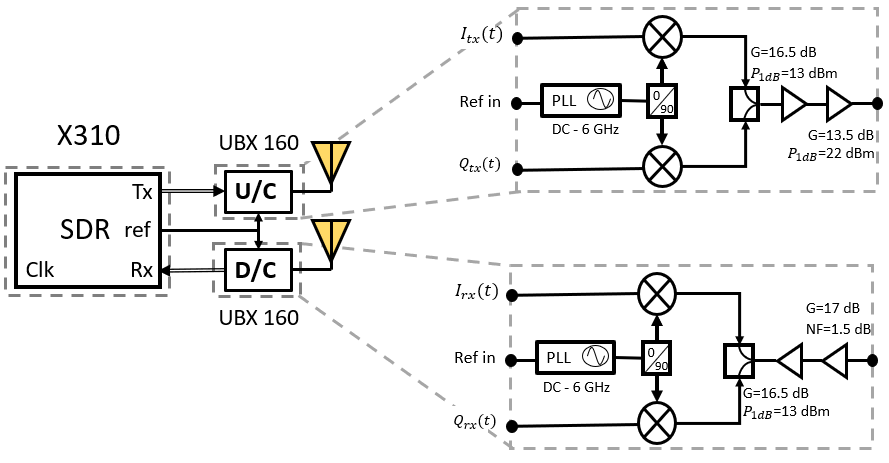}
  \caption{RF chain internal to each of the X310 SDRs.}
	\label{fig:X310}
\end{center}
\end{figure}

The architecture investigated in this work is based on a single primary node and multiple secondary nodes. As the objective is to demonstrate the ability to simultaneously measure internode range between multiple nodes with sufficient accuracy to support beamforming, we use continuous-wave transmitted signals, and we lock the reference oscillators via cable. Wireless frequency alignment can be implemented in various ways for a fully wireless system \cite{1408270,5482522,8889331}. 
%Open-loop CDA setups will contain a primary and one or more secondarys nodes. 
%The primary will set a reference coherent signal as well as a reference point in space for all the secondary nodes track and calibrate to. 
An image of the block diagram of the primary and secondary nodes can be seen in Fig. \ref{fig:nodes} (a) and (b) respectively. 
%The reference oscillators of all the secondarys were locked to the primary via SMA cables.
Each node consisted of two Ettus X310 SDRs, each of which were connected to one host computer running Windows 7 with 32 GB of RAM via 10 GB Ethernet cables. The X310s utilized two UBX 160 daughterboards which have operational bandwidths from DC to 6 GHz with an instantaneous bandwidth of up to 160 MHz. These daughterboards support complex up- and down-conversion for in-phase and quadrature mixing as well as internal amplification equivalent to 30 dB and 33.5 dB for the transmit and receive sides, respectively. A block diagram of the X310 RF chain can be seen in Fig. \ref{fig:X310}. The SDRs interfaced with the host computer using using LabVIEW 2018 where a maximum sampling rate of 25 MHz was possible which was limited by the data throughput between LabVIEW and the SDRs, restricting the maximum achievable instantaneous bandwidth to 12.5 MHz.

One SDR on each node transmitted the signal for distributed beamforming. The second SDR was used to either implement ranging to the primary node, or, on the primary node, to capture and retransmit any incoming ranging signals from the secondary nodes. Each secondary node transmits a version of the ranging waveform with a distinct stepped-frequency pattern. The primary node repeats any incoming signals in a continuous manner (i.e. no time scheduling was required). Each secondary node then processes the received signal via matched filter followed by the Kalman filter refinement step. Both of the two and three node experiments described below yielded SNR values of approximately 30 dB, which were determined using an eigenvalue decomposition approach\cite{6860836,8870258}. The secondary nodes then calculate the range, from which the relative phase of the beamforming carrier frequency was updated based on \eqref{phase}.

The beamforming signals were transmitted from each node at a carrier frequency of 1.5 GHz using 1.35-9.5 GHz ultra wideband log periodic antennas. Transmission of the ranging signals from the secondary nodes was implemented at a carrier frequency of 4.25 GHz and after reception at the mast node the ranging waveforms were retransmitted at a carrier frequency of 5.25 GHz, providing frequency diversity to mitigate crosstalk and multipath. The beamformed signals were captured on a Keysight MSO-X 92004A oscilloscope. The power levels of the individual signals were also recorded at each location by selectively turning on individual transmitters. 
%The local oscillators of the individual radios at each node were also connected via SMA cable.

\subsection{Two Node Beamforming}
Two experiments were conducted, one with two transmitting nodes and one with three transmitting nodes. In both cases the arrays were beamforming in an end-fire direction. The primary node was moved in both experiments, inducing relative phase errors between all nodes that was corrected via range estimation. An initial calibration procedure was implemented by adjusting the phases of the transmitted signals until maximum gain was obtained; after this point no adjustment was performed aside from that implemented by the ranging system. The calibration procedure effectively amounts to calibration of the phase delays present in each node, and can reasonably be implemented without monitoring the received power by using, e.g., power couplers at the transmitter outputs.

%A calibration procedure was performed where the received power of each of the nodes was measured on the scope and the output phase of each secondary node is tuned until that sum of the amplitudes from all the nodes in the array is seen at the receiver. The amplitudes are chosen so that the primary amplitude is roughly the sum of the secondary amplitudes. This is done so that when measurements without the phase locking procedure are taken, the resulting coherent amplitude will be roughly zero when the primary and secondary(s) are $180\degree$ out of phase.
\begin{figure} [t!]
	\centering
\includegraphics[width=\linewidth]{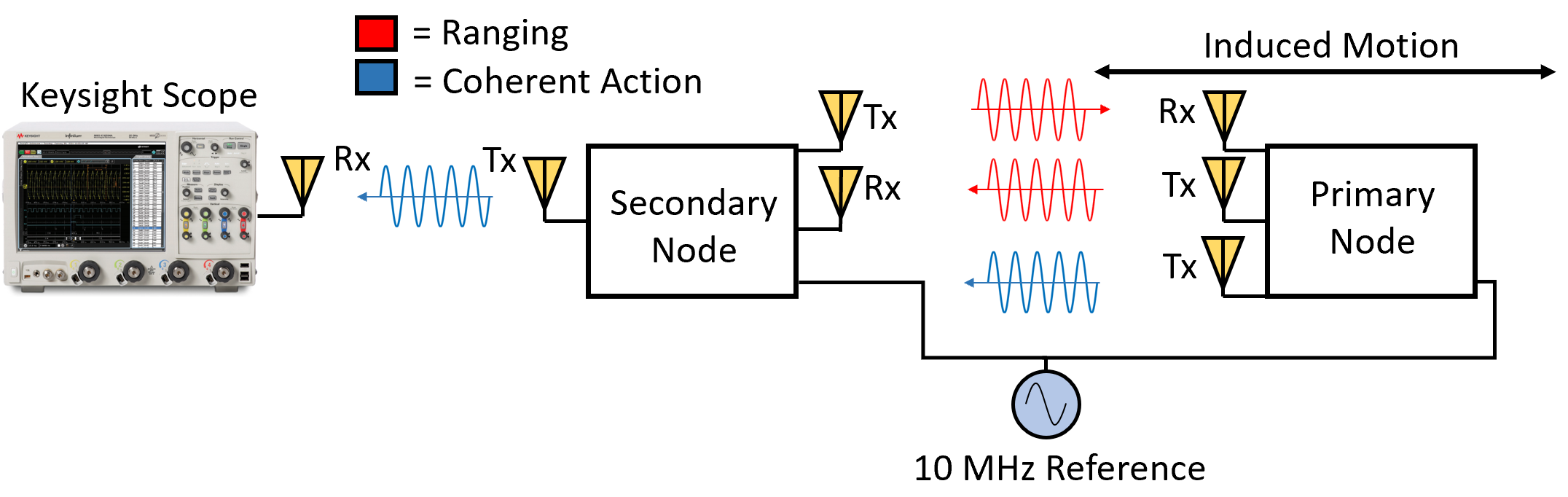}

(a)

\includegraphics[width=0.7\linewidth]{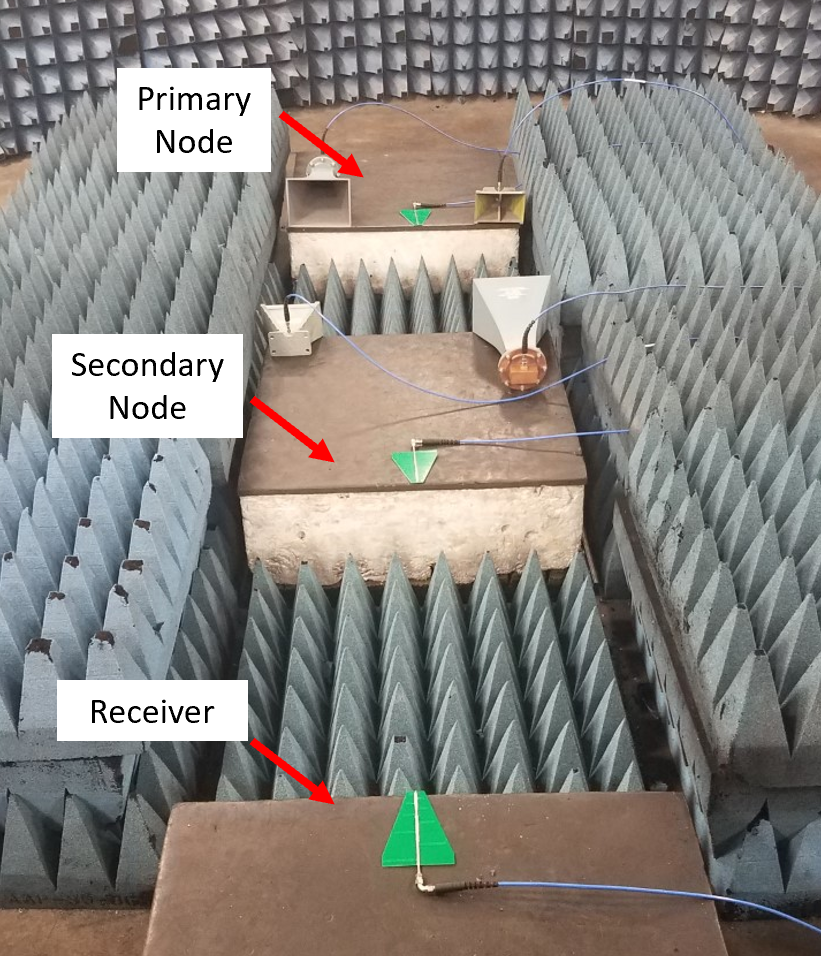}

(b)

\includegraphics[width=\linewidth]{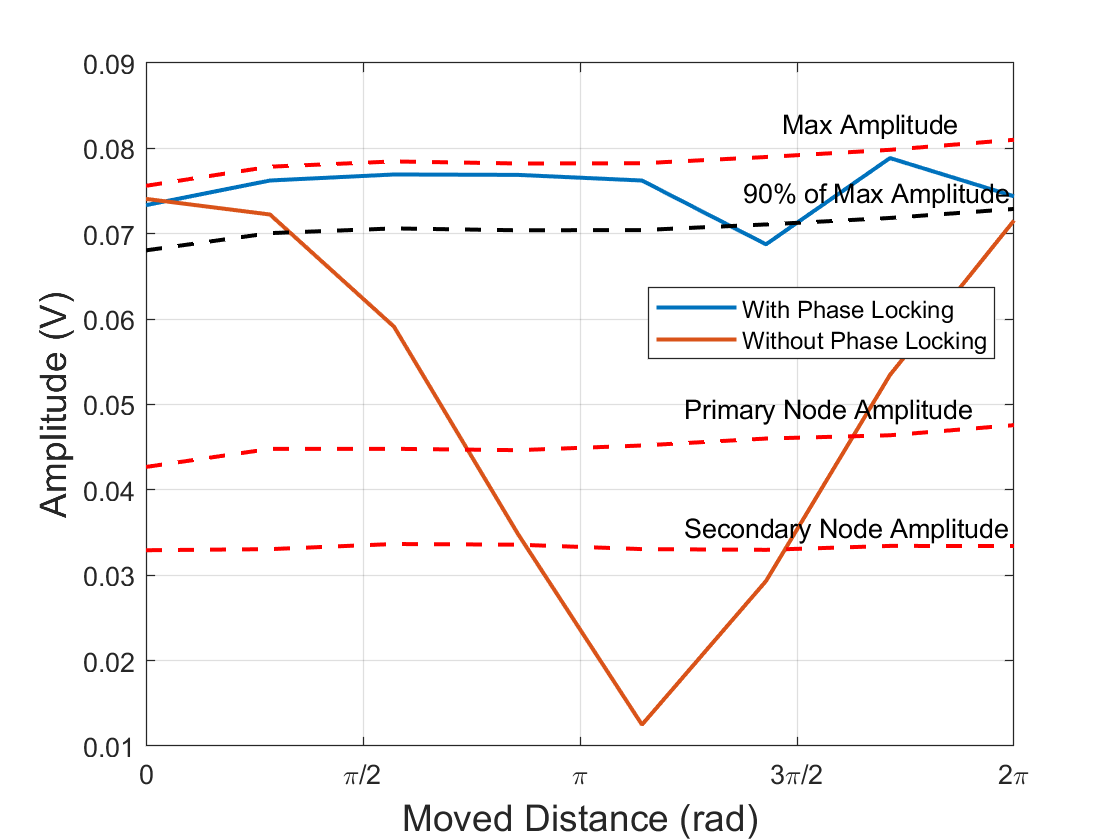}

(c)

	\caption{(a) Block diagram of the distributed two-node experiment. (b) Image of the experimental setup in the semi-enclosed arch range. (c) Measured results of coherent gain with and without performing range-based phase correction.}
	\label{fig:setup}
\end{figure}

The two-node distributed array consisted of a primary node and a single secondary node separated initially by a distance of 1.5 m. The receiving antenna at the destination was located at a distance of 1.5 m from the secondary node, and 3 m from the initial position of the primary node. The setup is shown in the diagram of Fig. \ref{fig:setup}(a), where the red colored signals indicate the ranging signals and the blue signals represent the beamforming signals. Fig. \ref{fig:setup}(b) shows the experimental system in a semi-anechoic chamber in the laboratory. The beamforming and ranging antennas were set up on blocks that were easily repositioned by sliding to different points. Absorber was used to mitigate reflections from nearby objects that were not fully shielded by absorber.
%This experiment consists of a primary node and one secondary node, separated by roughly 1.5 m, transmitting to a receiving antenna at about 1.5 m away from the secondary node in a linear fashion. This is to test the worst case scenario for coherent distributed arrays which happens when the array is oriented linearly and operating in endfire. As nodes in the array move the coherent signal phases will directly add in and out of phase while any angle off endfire will contain some residual signal proportional to $\sin(\theta)$ where $\theta$ is the steering angle.
%The local oscillators of the primary and secondary are connected via a SMA cable to ensure identical and error reference information is given to each node. The resulting coherent signal is captured on an Keysight MSO-X 92004A oscilloscope. 
The primary node was moved in approximately 2 cm increments towards the secondary to simulate motion of nodes in the array for a total of 20 cm, equal to the wavelength of the beamforming frequency. 
%This demonstrates the effects of moving form calibration, to 180\degree out of phase, back to in-phase but displaced in space by a wavelength. A block diagram of the system setup can be seen in Fig. \ref{fig:Setup} (a) where the red propagational symbols represent localization and the blue propagational symbols represent coherent action. An image of the distributed two node experimental setup in an partially enclosed arch range can be seen in Fig. \ref{fig:Setup} (b).
At each distance 100 snapshots containing 15 cycles of the 1.5 GHz signal were captured. The resulting 1,500 peak values were averaged to give the measured amplitude at each distance. 

Fig. \ref{fig:setup}(c) shows the measured results of the beamforming experiment. The red dashed lines show the individual amplitudes of the two transmitters, as well as the ideal summation of the two signal amplitudes, denoted max amplitude, which represents the maximum possible beamformed amplitude that would be achieved with error-free phase correction. The black dashed line indicates the $90\%$ coherent gain threshold. Two beamforming measurements are represented in the plot. The first is an uncorrected beamforming experiment where the ranging system was not utilized to update the transmission phase of the secondary node, represented by the orange curve. The signal begins initially at a high coherent amplitude value, before decreasing to a null and finally increasing again to a high amplitude value, clearly showing the constructive and destructive interference expected when no phase correction is implemented. The blue line shows the result of performing the same measurement with the ranging system automatically updating the phase of the beamforming signal. The beamformed amplitude is close to the ideal level for the majority of the test, indicating successful beamforming when the primary node was moved.

%The resulting accomplishments in amplitude power for the cases of performing the phase locking, not performing the phase locking, and individual nodes can be seen in Fig. \ref{fig:Setup} (c). From these results it can be seen that when the phase locking procedure is used, the coherent gain remains above 90\% over the total moved distance except one point where the it dips to 87\%. The case where phase locking is not used, the coherent gain drops to 15\% of the ideal coherent gain when the position of the platform has moved roughly half a wavelength of the coherent frequency.

\subsection{Three Node Experiment}

\begin{figure}[h!]
	\centering
\includegraphics[width=\linewidth]{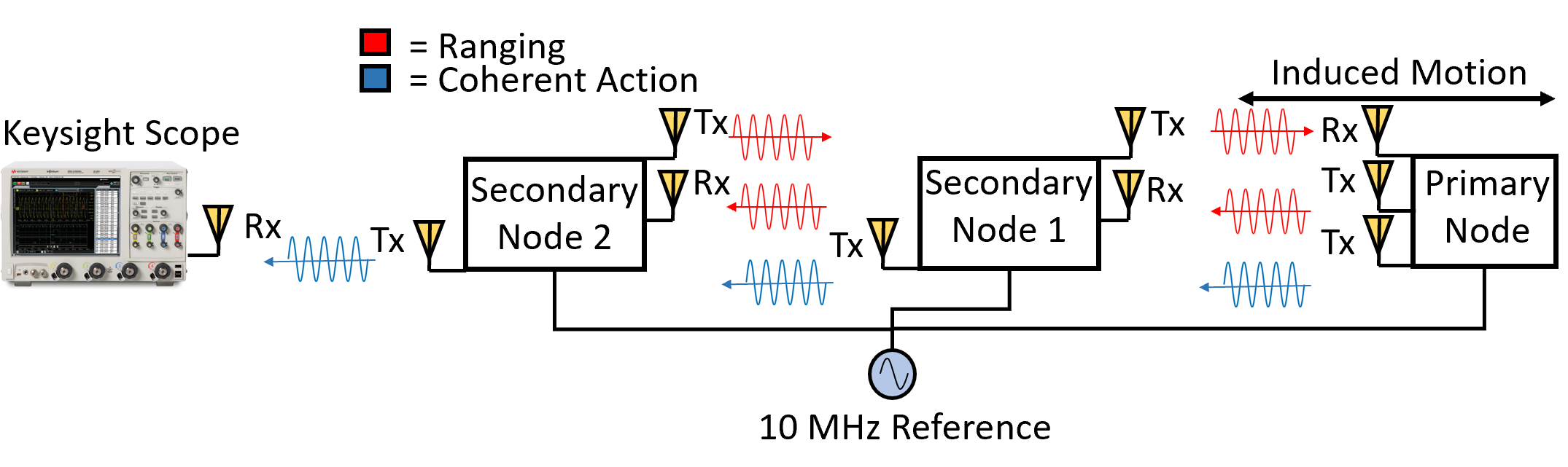}

(a)

\includegraphics[width=0.7\linewidth]{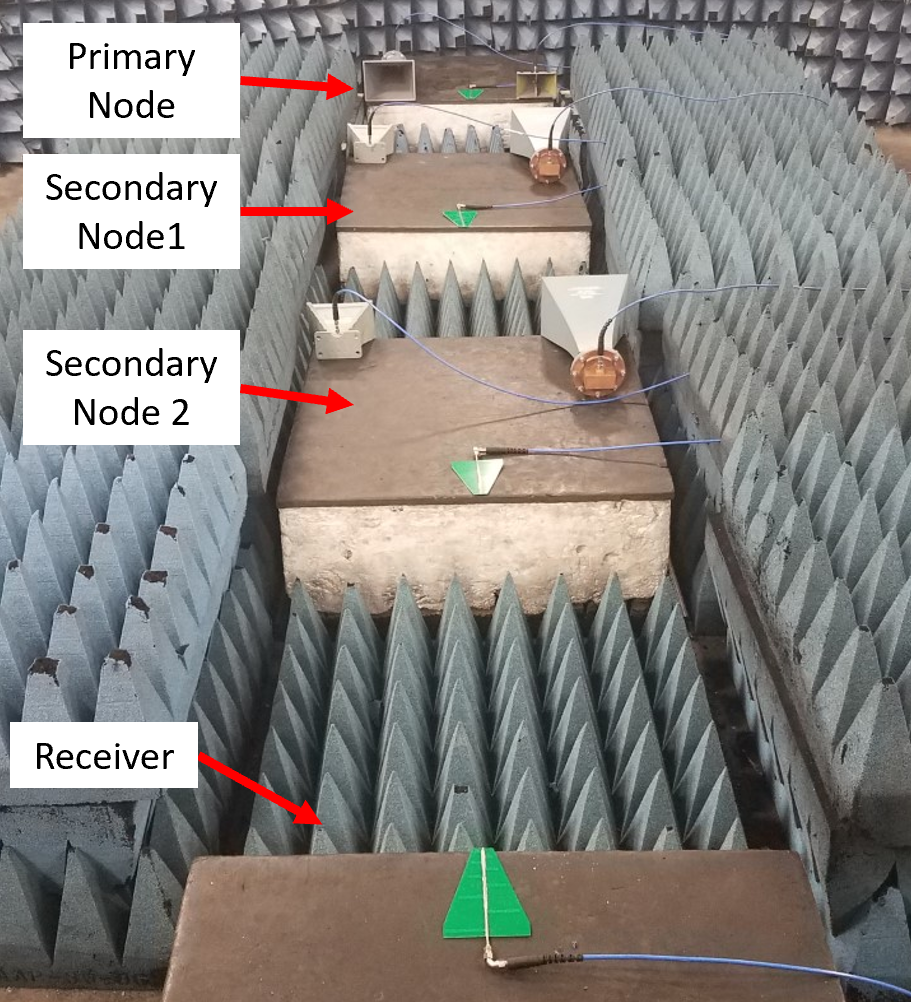}

(b)

\includegraphics[width=\linewidth]{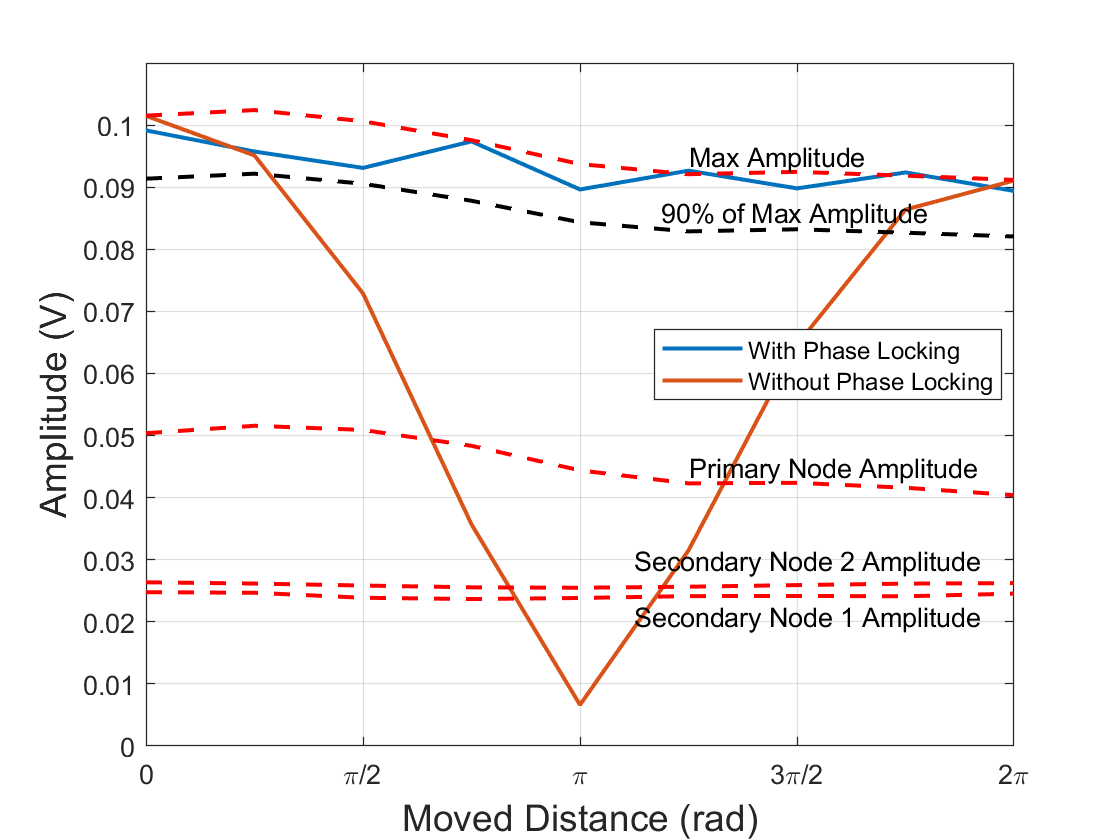}

(c)

	\caption{(a) Block diagram of the distributed three-node experiment. (b) Image of the experimental setup in the semi-enclosed arch range. (c) Measured results of coherent gain with and without performing range-based phase correction.}
	\label{fig:setup2}
\end{figure}

The two-node system was expanded upon to create a three-node open-loop beamforming system. To our knowledge, this was the first open-loop distributed beamforming demonstration with more than two transmitting nodes. The array was oriented in an end-fire configuration, again to demonstrate the most challenging beamforming case. The setup is shown in the diagram of Fig. \ref{fig:setup2}(a), where the red colored signals indicate the ranging signals and the blue signals represent the beamforming signals. Fig. \ref{fig:setup2}(b) shows the experimental system in a semi-anechoic chamber in the laboratory. The two secondary nodes were separated by 0.6 m, with first secondary node a distance of 1 m from the receiving antenna. The primary node started at a distance of 1 m from the second secondary node. The two secondary nodes were stationary, and transmitted signals to the primary node, which captured and retransmitted the signals. No time scheduling was used for the two secondary nodes; the stepped-frequency waveform supports simultaneous multinode operation thus no scheduling was required. The primary node was moved in 2m increments for a total range of 20 cm. In the same way as the two-node experiment, at each distance 100 snapshots containing fifteen cycles of the 1.5 GHz signal were captured. The resulting 1,500 peak values were again averaged to give the measured amplitude at each distance.

The measured results of the three-node experiment are shown in Fig. \ref{fig:setup2}(c). Here the two secondary node individual powers were again constant since the secondary nodes were stationary. The uncorrected beamforming measurement showed a clear null, dropping to 7\% of the available total power when the primary was in a location causing destructive interference. When the ranging system was utilized, the system maintained a high-gain beamforming signal, achieving a coherent power level consistently above the 90\% threshold.

%The three node experiment has a very similar to the two node setup except there is a second secondary node. This node is added linearly between the primary and receiver. An image of the block diagram of the three node setup can be seen in Fig. \ref{fig:setup2} (a). The receiver was place approximately 1 m away from secondary 2. Secondary 1 and 2 were separated by roughly 2 ft (0.6 m) the width of the box the antennas were placed on. The primary node was placed about 1 m away from secondary 1. The experimental setup in the semi-enclosed arch range can be seen in Fig. \ref{fig:Setup2} (b). The measurement was again taken where the primary node is moved towards the two secondary nodes in 2 cm increments and the resulting coherent signal is received by the scope. The scope again takes 100 captures of 15 cycles of the 1.5 GHz signal where the resulting 1,500 amplitudes are averaged at every distance. An image of the measured results for each distance can be seen in Fig. \ref{fig:Setup2} (c).  Again it can be seen form this result that when the phase correction is being applied all of the measured points remain above 90\% coherent gain and when the correction is not being applied the coherent gain drops as low as 7\% and is again extremely position dependent.
%%%%%%%%%%%%%%%%%%%%%%%%%%%%%%%%%%%%%%%%%%%%%%%%%%%%%%%%%%%%%%%%%%%%%
\section{Conclusion}\label{Sec6}
Distributed beamforming in open-loop systems requires scalable, high-accuracy coordination. Localization represents one of the fundamental coordination tasks necessary for open-loop beamforming, and is arguably the most challenging considering the error tolerances for microwave and millimeter-wave systems. The approach demonstrated in this work utilizes time of flight estimations whose accuracy was maximized by a one dimensional Kalman filter to enable real time dynamic phase adjustments for mobile open-loop beamforming arrays. This method provides a solution for scalable, high-accuracy localization that can support distributed beamforming between multiple nodes at transmission frequencies relevant for numerous wireless applications. Combined with wireless frequency synchronization and time alignment, fully open-loop distributed beamforming will be possible on future distributed wireless systems.

%In conclusion, to enable successful open-loop beamforming, relative positioning needs to be know within a $\lambda/15$ of the operational frequency to ensure that the phase adjustments of array dynamics maintain at least 90\% coherent gain at the receive location. In this work we have demonstrated a 1.5 GHz operational frequency with dynamic phase adjustments form ranging data provided by delay estimations. For operations in the microwave range, estimations require sub-cm accuracies provided by a TTSFW with 12.5 MHz total bandwidth with a variety of pulses enabling various array sizes. The delay estimation provided from this waveform is converted to a phase shift equal to the relative distance between the primary and corresponding secondary node. Operations containing a primary node and up to two secondary nodes simultaneously performing their relative phase adjustments have been shown. For both the experiments, two and three node, the phase adjustment procedure proved to be effective at maintaining coherent gain, effectively making the array independent of node motion. When no phase adjustment is being performed the array gain becomes extremely dependent on array motion.
%%%%%%%%%%%%%%%%%%%%%%%%%%%%%%%%%%%%%%%%%%%%%%%%%%%%%%%%%%%%%%%%%%%%%
\bibliographystyle{IEEEtran}
\bibliography{IEEEabrv,Distributed_array_arXiv_submission_-_Final}

\end{document}